\documentclass[twocolumn,showpacs,aps,floatfix,superscriptaddress]{revtex4}
\usepackage{amsmath}
\usepackage{graphicx}
\begin{document}
\title{On The Structure of Competitive Societies}
\author{E.~Ben-Naim}\email{ebn@lanl.gov}
\affiliation{Theoretical Division and Center for Nonlinear Studies,
Los Alamos National Laboratory, Los Alamos, New Mexico 87545}
\author{F.~Vazquez}\email{fvazquez@buphy.bu.edu}
\affiliation{Theoretical Division and Center for Nonlinear Studies,
Los Alamos National Laboratory, Los Alamos, New Mexico 87545}
\affiliation{Department of Physics, Boston University, Boston,
Massachusetts, 02215 USA}
\author{S.~Redner}\email{redner@bu.edu}
\affiliation{Theoretical Division and Center for Nonlinear Studies,
Los Alamos National Laboratory, Los Alamos, New Mexico 87545}
\affiliation{Department of Physics, Boston University, Boston,
Massachusetts, 02215 USA}
\begin{abstract}
  We model the dynamics of social structure by a simple interacting particle
  system.  The social standing of an individual agent is represented by an
  integer-valued fitness that changes via two offsetting processes.  When two
  agents interact one advances: the fitter with probability $p$ and the less
  fit with probability $1-p$.  The fitness of an agent may also decline with
  rate $r$.  From a scaling analysis of the underlying master equations for
  the fitness distribution of the population, we find four distinct social
  structures as a function of the governing parameters $p$ and $r$.  These
  include: (i) a static lower-class society where all agents have finite
  fitness; (ii) an upwardly-mobile middle-class society; (iii) a hierarchical
  society where a finite fraction of the population belongs to a middle class
  and a complementary fraction to the lower class; (iv) an egalitarian
  society where all agents are upwardly mobile and have nearly the same
  fitness.  We determine the basic features of the fitness distributions in
  these four phases.

\end{abstract}
\maketitle

\section{Introduction}

The emergence of class structure in society is a ubiquitous phenomenon in the
biological and the social sciences \cite{idc,rvg,hgl,eow}.  Social
hierarchies have been widely observed in animal populations including insects
\cite{eow1}, mammals \cite{wca,amg,sf}, and primates \cite{vs}, as well as
human communities \cite{idc1}.

The possibility of quantitative modeling of social phenomena using concepts
and techniques borrowed from the physical sciences is rapidly gaining
appreciation.  Examples of such modeling include the wealth distribution
\cite{ikr,dy}, opinion dynamics \cite{wdan,bkr,smo}, and rumor propagation
\cite{wf,gn}.  Such approaches typically draw analogies between individual
agents in the social system and particles in a corresponding physical system
and then identifying macroscopically observed phenomena with microscopic
agent-agent interactions \cite{ww,vfh,ckfl}.

In this spirit, we seek to tie the emergence of social structures to
specific interactions between agents within a general version of the
recently-introduced advancement-decline process \cite{btd,SS,br}.  In
our model, the social standing of each agent is characterized by a
single number, its fitness.  Agents increase their fitness by
interacting with other agents and also, their fitness may decline
spontaneously.  This simple model has only two parameters: the
probability that the fitter agent advances in an interaction and the
rate of decline.

We find that a rich variety of familiar social structures emerges as a result
of the competition between advancement and decline.  When decline dominates,
the society is static and the fitness distribution approaches a steady state.
When the decline rate is comparable to the advancement rate, the society is
dynamic and the characteristic fitness of the population increases linearly
with time.  In this case, there are several possibilities.  When the less fit
agent benefits from social interactions, an egalitarian society arises in
which all agents advance at the same rate.  Consequently, inequalities among
agents are small.  On the other hand, when the fitter agent tends to benefit
in competitions, agents advance at different rates and social inequalities
increase with time.  Depending on the relative influence of advancement and
decline, either the entire population or only a fraction of it may be
upwardly mobile.  In the latter case, the society consists of a static lower
class and an upwardly-mobile middle class.

In Section~\ref{ADP}, we introduce the general advancement-decline process
and the governing master equations.  The overall class structure and the
statistics of the mobile middle class are obtained using scaling analysis in
Section~\ref{mobility}.  The basic features of the egalitarian society are
investigated in section~\ref{equality}, where the cumulative fitness
distribution may be largely determined by linear traveling wave analysis.  In
Section~\ref{poverty}, the statistics of the lower class, where the fitness
distribution is steady, are determined.  We conclude in section~\ref{conc}.

\section{The Advancement-Decline Model}
\label{ADP}

\begin{figure}[b]
\vspace*{0.cm}
\centerline{\includegraphics*[width=0.22\textwidth]{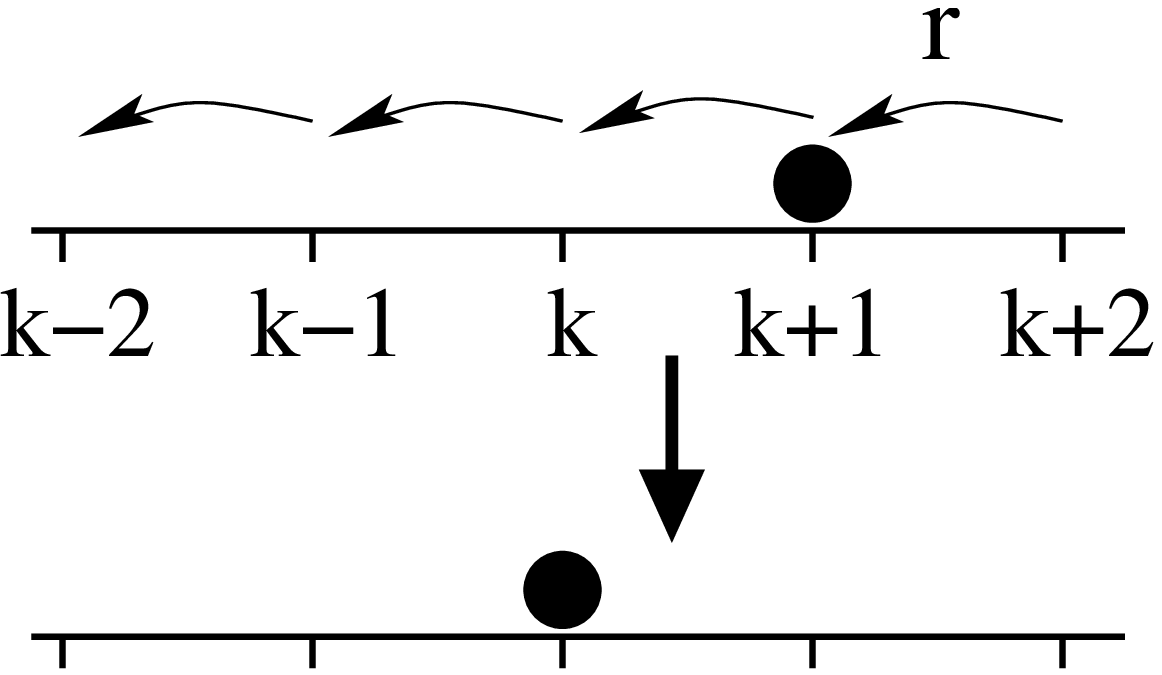}}
\vskip 0.3in
\includegraphics*[width=0.4\textwidth]{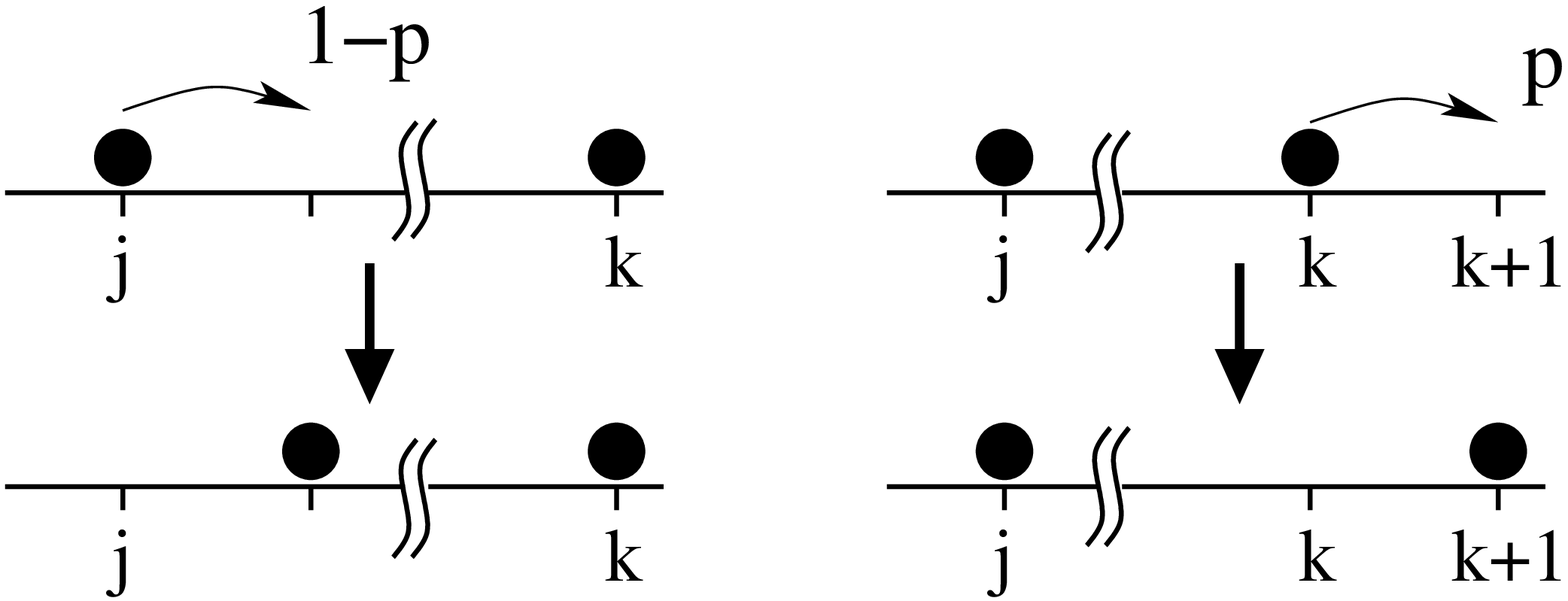}
\caption{The elemental processes of the advancement-decline model.  Top: the
  decline process.  Bottom: the advancement process, with either the fitter
  agent advancing (right) or the less fit agent advancing (left).}
\label{processes}
\end{figure}

We model a scenario in which the social status of an agent benefits from
increased social interactions, while solitude or isolation have the opposite
effect.  Indeed, highly connected individuals often have better access to
information, resources, and power, that are often gained as a result of
social interactions.  Thus, in our model there are two competing evolutionary
processes that influence the fitness of agents: (i) advancement via social
interactions, and (ii) decline due to the lack of interactions
(Fig.~\ref{processes}).  For simplicity, social standing is represented by a
single parameter, the integer-valued fitness $k\geq 0$.

\medskip\noindent{\it (i) Advancement.} Agents interact in pairs, and as a
result of the interaction, only one advances.  There are two
possibilities: either the fitter agent advances or the less fit
advances.  We allow the fitter agent to advance with probability $p$
and the less fit agent to advance with probability $1-p$.  Thus,
when two agents with fitness $k$ and fitness $j$ interact (with $k>
j$), the outcome is
\begin{eqnarray*}
\label{up}
(k,j)&\to& (k+1,j) \quad{\rm with\ probability~} p,\\
(k,j)&\to& (k,j+1) \quad{\rm with\ probability~} 1-p.
\end{eqnarray*}
For $p=1$ the fitter agent always advances \cite{br}, while for $p=0$
the less fit agent always advances.  The interaction rules are defined
so that one randomly-chosen agent advances when two equally-fit agents
interact.  Without loss of generality, the interaction rate is set to
$1/2$.  Also, we consider the thermodynamic limit where the number of
agents is infinite.

\medskip\noindent{\it (ii) Decline.} In the decline step, the fitness of an
individual decreases according to
\begin{eqnarray*}
\label{down}
k\to k-1
\end{eqnarray*}
with rate $r$.  This process reflects a natural tendency for fitness to
decrease in the absence of social activity.  We impose the lower limit for
fitness to be $k=0$; once an individual reaches zero fitness, there is no
further decline.

Our goal is to understand how the fitness distribution of a population
evolves as a function of the two model parameters, the advancement
probability $p$ and the decline rate $r$.  Let $f_k(t)$ be the fraction of
agents with fitness $k$ at time $t$.  In the mean-field limit, where any pair
of agents is equally likely to interact, the fitness distribution obeys the
master equation
\begin{equation}
\label{dis-eq}
\begin{split}
\frac{df_k}{dt}&=r(f_{k+1}-f_k)+p(f_{k-1}F_{k-1}-f_kF_k)\\
&+(1-p)(f_{k-1}G_{k-1}-f_kG_k)+\frac{1}{2}(f_{k-1}^2-f_k^2)\,.
\end{split}
\end{equation}
Here $F_k=\sum_{j=0}^{k-1} f_j$ and $G_k=\sum_{j=k+1}^\infty f_j$
are the respective cumulative distributions of agents with fitness
less than $k$ and fitness greater than $k$.  The boundary condition
is $f_{-1}(t)=0$. The first pair of terms accounts for decline, the
second pair of terms describes interactions where the stronger agent
advances, and the third pair of terms accounts for interactions
where the weaker agent advances. The last pair of terms describes
interactions between two equal agents and it reflects that when two
such agents interact, only one of them advances. The prefactor $1/2$
arises because there are half as many ways to chose equal agents as
there are for different agents. We consider the initial condition
where all agents have the minimal fitness $f_k(0)=\delta_{k,0}$.

It proves useful to rewrite the evolution equation in a closed form that
involves only the cumulative distribution.  Summing the rate equations
(\ref{dis-eq}) and using the relations $f_k=F_{k+1}-F_k$ and $G_k=1-F_{k+1}$,
the cumulative distribution $F_k$ obeys
\begin{eqnarray}
\label{cum-eq1}
\begin{split}
\frac{dF_k}{dt}&
=r(F_{k+1}-F_k)+pF_{k-1}(F_{k-1}-F_k)\\
&+(1-p)(1-F_k)(F_{k-1}-F_k)-\frac{1}{2}(F_k-F_{k-1})^2.
\end{split}
\end{eqnarray}
The boundary conditions are $F_0=0$, $F_\infty=1$, and the initial
condition is $F_k(0)=1$ for $k\geq 1$. There is a one-to-one
correspondence between the four terms in equations (\ref{dis-eq}) and
(\ref{cum-eq1}). The master equation for the cumulative distribution
can be simplified by consolidating the advancement terms
\begin{equation}
\label{cum-eq}
\begin{split}
\frac{dF_k}{dt}
&=r(F_{k+1}-F_k)+(1-p)(F_{k-1}-F_k)\\
&+(p-1/2)\left(F^2_{k-1}-F^2_k\right).
\end{split}
\end{equation}

The mean fitness $\langle k\rangle =\sum_k k f_k$ evolves according to
\begin{equation}
\label{kdot}
\frac{d\langle k\rangle}{dt} =\frac{1}{2}-r(1-f_0)
\end{equation}
This result can be derived directly by summing the master equations
(\ref{dis-eq}) or even simpler, from the definition of the
advancement-decline process.  The first term accounts for advancement,
where interactions occur with rate $1/2$ such that each interaction
advances only one agent.  The second term stems from decline and
reflects the fact that all agents except for the least-fit ones
decline with rate $r$.

We now discuss the basic social structures that emerge from the
solution to the master equation.

\section{Emergence of social structures}
\label{mobility}

\subsection{Scaling solution}

We determine the class structure of the population via a simple scaling
analysis of the master equation.  Let us take the continuum limit of the
master equation by replacing differences with derivatives, $F_{k+1}-F_k\to
\partial F/\partial k$. To first order in this ``spatial'' derivative, we
obtain the nonlinear partial differential equation
\begin{equation}
\label{cum-eq-cont}
\frac{\partial F}{\partial t}=\left[p+r-1-(2p-1)F\right]
\frac{\partial F}{\partial k}\,.
\end{equation}
When the spatial derivative and the temporal derivative balance, the typical
fitness increases linearly with time, $k\sim t$. Therefore, we make the
scaling ansatz
\begin{equation}
\label{scaling-form}
F_k(t)\simeq \Phi\left(\frac{k}{t}\right).
\end{equation}
The boundary conditions are $\Phi(0)=0$ and $\Phi(\infty)=1$.

Substituting this scaling form in Eq.~(\ref{cum-eq-cont}), the
partial-differential equation reduces to the ordinary differential equation
\begin{equation}
\label{scaling-eq}
\left[(p+r-1+x)-(2p-1)\Phi(x)\right]\frac{d\Phi}{dx}=0,
\end{equation}
where the prime denotes differentiation with respect to the scaling variable
$x\equiv k/t$.  The solution is either $d\Phi/dx=0$, {\it i.e.},
\begin{equation}
\label{scaling-sol-const}
\Phi(x)={\rm constant},
\end{equation}
or the linear function
\begin{equation}
\label{scaling-sol-lin}
\Phi(x)=\frac{p+r-1}{2p-1}+\frac{x}{2p-1}\,.
\end{equation}
Using these two solutions and invoking (i) the boundary conditions
$\Phi(0)=0$ and $\Phi(\infty)=1$, (ii) the bounds $0<\Phi(x)<1$, (iii)
monotonicity of the cumulative distribution, $d\Phi(x)/dx\geq 0$, and
(iv) the assumption that the scaling function changes continuously
with $p$ and $r$, we can then deduce the four possible social
structures of the population.

\begin{figure}[t]
\vspace*{0.cm} \includegraphics*[width=0.35\textwidth]{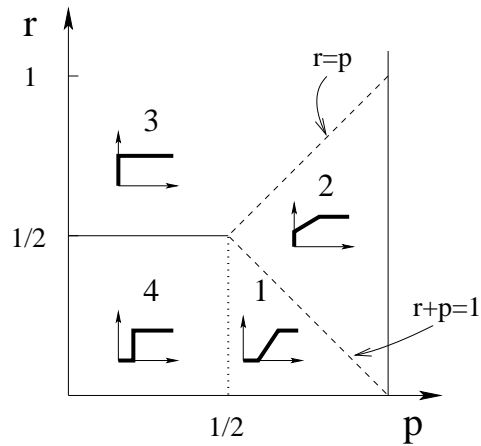}
\caption{Phase diagram of the advancement-decline model.  The small graphs in
  each region are sketches of the scaled cumulative fitness distribution.}
\label{classes}
\end{figure}

\begin{figure}[t]
 \vspace*{0.cm}
\includegraphics*[width=0.4\textwidth]{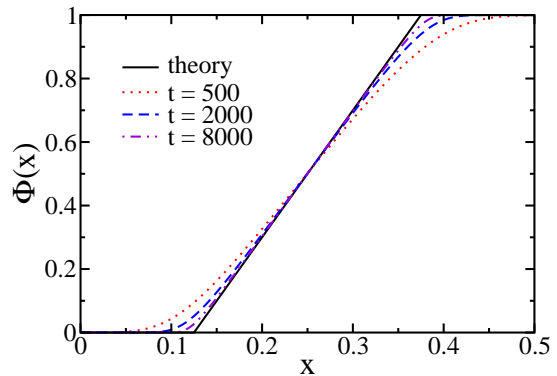}
\caption{Middle-Class society. The scaled cumulative fitness distribution
  $\Phi(x)$ versus $x=k/t$ at different times for $p=5/8$ and $r=1/4$.}
\label{fig-phi-mc}
\end{figure}

\medskip \noindent{\it 1. Middle-Class Society}: We first examine the
conditions for the linear scaling function (\ref{scaling-sol-lin}) to apply.
First, the cumulative scaling function (\ref{scaling-form}) must be a
monotonically increasing function.  Therefore, the linear solution
(\ref{scaling-sol-lin}) holds only when its slope is positive, that is, when
\hbox{$p>1/2$}. Second, the scaling function is bounded, \hbox{$0\leq
  \Phi(x)\leq 1$}; this condition implies the lower and upper bounds
\begin{eqnarray}
x_-=1-(p+r) \qquad{\rm and}\qquad
x_+=p-r
\end{eqnarray}
on the scaled fitness.  The obvious constraints $x_->0$ and $x_+>0$
lead to the conditions $p+r<1$ and $p>r$. By imposing continuity, as
well as the limiting behaviors $\Phi(x)=0$ and $\Phi(x)=1$ outside the
linear region, the scaled cumulative distribution is the piecewise
linear function (Fig.~\ref{fig-phi-mc}):
\begin{equation}
\label{phi-mc}
\Phi_{\rm M}(x)=
\begin{cases}
0& 0<x<x_-\\
{\displaystyle \frac{p+r-1}{2p-1}+\frac{x}{2p-1}}&x_-<x<x_+\\
1& x_+<x.
\end{cases}
\end{equation}
This behavior describes a middle class society where {\em all\/}
agents are upwardly mobile, as their fitness improves linearly with
time. In this case, social inequalities also increase indefinitely
with time: the agents at the bottom of the middle class have fitness
$k_-=[1-(p+r)]t$ and the richest agents have fitness $k_+=(p-r)t$.
The middle-class society lies within the triangular region defined by
the lines $r+p=1$, $r=0$, and $p=1/2$, shown in Fig.~\ref{classes}.

\begin{figure}[t]
\includegraphics*[width=0.4\textwidth]{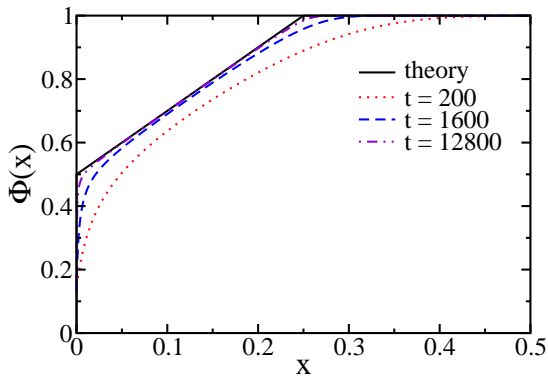}
\caption{Hierarchical society. The scaled cumulative fitness
distribution $\Phi(x)$ versus $x=k/t$ at different times for $p=3/4$
and $r=1/2$.}
\label{fig-phi-hs}
\end{figure}

\medskip\noindent{\it 2. Hierarchical Society}: Along the line $r+p=1$, the
fitness of the poorest agents vanishes. Moreover, the linear scaling solution
(\ref{scaling-sol-lin}) has a finite positive value at zero fitness for a
range of parameter values $p$ and $r$.  These two observations suggest the
existence of another type of piecewise linear solution with $\Phi(0)>0$.  The
bounds $0<\Phi(0)<1$ impose the conditions $p+r>1$ and $r>p$. In this region,
the scaling function has two distinct components (Fig.~\ref{fig-phi-hs})
\begin{equation}
\label{phi-hs}
\Phi_{\rm H}(x)=
\begin{cases}
{\displaystyle \frac{p+r-1}{2p-1}+\frac{x}{2p-1}}&0<x<x_+\\
1& x_+<x.
\end{cases}
\end{equation}
Thus, we find a hierarchical society (Fig.~\ref{classes}) that includes both an
upwardly-mobile middle class and a static lower class.  The lower class
consists of a finite fraction
\begin{equation}
\label{l}
L=\frac{p+r-1}{2p-1}
\end{equation}
of agents with zero fitness (in scaled units).  In section~\ref{poverty}, we
examine the lower class more closely and show that its fitness distribution
is time-independent and extends only over a finite range.

\medskip\noindent{\it 3. Lower-Class Society}: When the fraction $L$ of
agents with zero fitness reaches 1, the entire population is poor.  For
$p>1/2$, the condition $L=1$ occurs on the boundary $p=r$.  At this point the
fitness distribution becomes a step function,
\begin{equation}
\label{phi-lc}
\Phi_{\rm L}(x)=\Theta(x),
\end{equation}
with $\Theta(x)=0$ for $x\leq 0$ and $\Theta(x)=1$ for $x>0$.  We
therefore conclude that there is a region of the phase diagram where
the scaled fitness of the entire population is zero.  For any initial
state, the fitness distribution quickly approaches the step-function
in a lower-class society.

\medskip\noindent{\it 4. Egalitarian Society}: There is another
region of the phase diagram where the fitness distribution also
becomes a step function.  When $p=1/2$ and $r<1/2$, then both $x_+$
and $x_-$ are equal to $1/2-r$.  Therefore
\hbox{$\Phi(x)=\Theta(x-v)$} with $v=1/2-r$. Since the scaling
function must change in a continuous fashion, we conclude that for
$p<1/2$, the scaling function is again a step function but with a jump
at non-zero fitness. That is
\begin{equation}
\label{phi-es}
\Phi_{\rm E}(x)=\Theta(x-v).
\end{equation}
In this egalitarian society, all agents have the same scaled fitness
$x=v$ or alternatively the fitness $k\approx vt$.  The velocity $v$
follows easily from the average fitness (\ref{kdot}). Since all agents
advance at constant rate, then the term $-rf_0$ is negligible and
therefore, the propagation velocity is
\begin{equation}
\label{v}
v=\frac{1}{2}-r.
\end{equation}
In section \ref{equality}, we show that in this society, the fitness
differences between agents are small and do not grow with time.  This
is the sense in which the society is egalitarian.  When $p<1/2$, the
weaker agent preferentially benefits in an interaction, so that the
rich effectively supports the poor.  We also note that the lower class
and the egalitarian society share one common feature: they do not have
a middle class.  The boundary between these two phases, determined by
the condition $v=0$, is the line $r=1/2$ (Fig.~\ref{classes}).

Our numerical integration of the evolution equations confirms the
overall picture of four different social structures
(Fig.~\ref{classes}): a middle class society (Fig.~\ref{fig-phi-mc}),
a hierarchical society (Fig.~\ref{fig-phi-hs}), a lower-class society
as in (\ref{phi-lc}), and an egalitarian society as in (\ref{phi-es}).
The numerical data was obtained by integrating $F_k$ for $0\leq
k<10000$ using a fourth-order Adams-Bashforth method \cite{dz}, with a
specified accuracy of $10^{-10}$ in the distribution $F_k$.

\subsection{Refinements to the Scaling Solutions}

Our numerical results for the cumulative distribution $F_k$, when
plotted versus the scaling variable $x=k/t$, smoothly approaches the
appropriate expressions for the piecewise linear scaling function
$\Phi(x)$ derived in the previous subsection (Figs.~\ref{fig-phi-mc}
\& \ref{fig-phi-hs}).  As time increases, the fitness distribution
narrows. The simulations also show that the approach to the scaling
solution is slowest in the vicinity of the extremes of the middle
class $x=x_-$ and $x=x_+$ (with $x_-=0$ for the hierarchical phase).

The correction to scaling near these extrema can be determined by
keeping derivatives up to second order in the continuum limit of the
master equation.  This approximation gives the nonlinear diffusion
equation \cite{gbw,jmb}
\begin{eqnarray}
\label{cum-eq-second}
\frac{\partial F}{\partial t}&=&(r+p-1)
\frac{\partial F}{\partial k}+
\frac{1}{2}(1+r-p)\frac{\partial^2 F}{\partial^2 k}\\
&+&(1-2p)F\left[\frac{\partial F}{\partial k}
-\frac{1}{2}\frac{\partial^2 F}{\partial^2 k}\right]
+(p-1/2)\left(\frac{\partial F}{\partial k}\right)^2.\nonumber
\end{eqnarray}
The linear terms are separately displayed in the first line and the nonlinear
terms in the second.

Let us first consider the poorest agents, {\it i.e}, the behavior close to
$x=x_-$. Since the cumulative fitness distribution is small near this point,
the nonlinear terms can be neglected and the governing equation
(\ref{cum-eq-second}) reduces to the standard convection-diffusion equation
\begin{equation}
\label{de}
\frac{\partial F}{\partial t}+v_-\,
\frac{\partial F}{\partial k}=D_-\,\frac{\partial^2 F}{\partial^2 k}
\end{equation}
with propagation velocity $v_-=x_-=1-p-r$ and diffusion coefficient
$D_-=(1-p+r)/2$.  Indeed, since the fitness distribution is obtained from the
cumulative distribution by differentiation, $f=\partial F/\partial k$, the
fitness distribution satisfies the same equation (\ref{cum-eq-cont}) as the
cumulative distribution.

For the middle-class society, we therefore conclude that the bottom of
the middle class has a Gaussian tail, with the center of the Gaussian
located at $k_-=v_-t$ and with width $\sqrt{D_-t}$.  The same analysis
can be carried out for the hierarchical society, where the quantity
$F-L$ now satisfies the diffusion equation with zero velocity $v_-=0$
and diffusivity $D_-=r$.  Conversely, the distribution for the top end
of the middle class can be obtained by analyzing $1-F$.  It is
immediate to show that this quantity again obeys Eq.~(\ref{de}) with
velocity $v_+=x_+=p-r$ and diffusivity $D_+=(r+p)/2$.  We conclude
that the extremes of the middle class are characterized by Gaussian
tails whose extents grow diffusively with time.  In terms of the
scaling variable $x$, the deviation from the scaling function
$\Phi(x)$ is appreciable only within a region of whose width is
shrinking as $t^{-1/2}$.

For the special case $p=1/2$, the nonlinear terms vanish and the
fitness distribution is described {\it exactly} by the linear
convection-diffusion equation (\ref{de}) with drift velocity $v=1/2-r$
and diffusion coefficient $D=(r+p)/2$ (the nonlinear term is
negligible).  Thus there is a drift toward smaller fitness for $r>1/2$
and the fitness distribution approaches a steady-state profile that
decays exponentially with fitness.  In the opposite case of $v>0$, the
fitness distribution is simply a Gaussian that drifts to larger
fitness with velocity $\frac{1}{2}-r$ and whose width is proportional
to $\sqrt{Dt}$.  In the case of $p=1/2$, the relative position of an
agent in the society is irrelevant and advancement reduces to a pure
random walk \cite{bkm}.

\section{Egalitarian Society}
\label{equality}

\begin{figure}[t]
 \vspace*{0.cm}
\includegraphics*[width=0.4\textwidth]{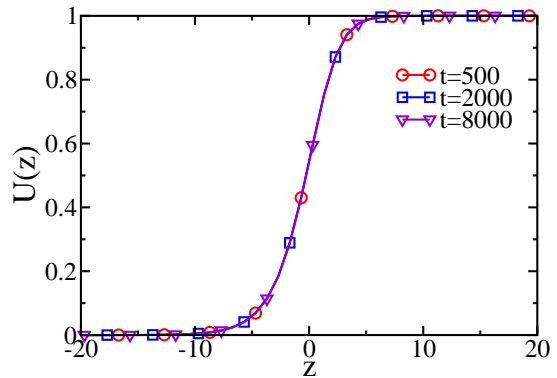}
\caption{The scaled cumulative fitness distribution $U(z)$ versus
$z=k-vt$, where $v=1/2-r$ is the speed of wavefront at different
times for $p=1/4$ and $r=1/4$ (egalitarian society).
\label{fig-wave}}
\end{figure}

In the egalitarian phase, the step function form of the scaling solution,
Eq.~(\ref{phi-es}), suggests that the fitness distribution has the traveling
wave form
\begin{equation}
\label{wave}
F_k(t)\to U(k-vt)
\end{equation}
with the propagation velocity (\ref{v}).  This is confirmed by
numerical integration of the master equation (\ref{cum-eq}), as shown
in Fig.~(\ref{fig-wave}).

To determine the shape of the wave $U(z)$ analytically, we substitute
the waveform (\ref{wave}) into the master equation (\ref{cum-eq}) to
give the nonlinear difference-differential equation for $U(z)$
\begin{eqnarray}
\label{wave-eq}
-vU'(z)\!
&=&\!r[U(z\!+\!1)\!-\!U(z)]\!+
\!(1\!-\!p)[U(z\!-\!1\!)-\!U(z)]\nonumber\\
&+&(p-1/2)[U^2(z-1)-U^2(z)].
\end{eqnarray}
The boundary conditions are $U(-\infty)=0$ and $U(\infty)=1$.

\subsection{Waveforms in the tail regions}

We apply standard linear analysis in the tail regions to deduce the
leading and trailing shapes of the waveform.  When $z\to-\infty$, then
$U\ll 1$ and therefore $U^2\ll U$.  To first order in $U$,
Eq.~(\ref{wave-eq}) becomes
\begin{eqnarray*}
vU'+r[U(z\!+\!1)-U(z)]+(1\!-\!p)[U(z\!-\!1)-U(z)]=0.
\end{eqnarray*}
The behavior in this case is determined by the balance between decline
and advancement events where the less fit agent advances.  The
solution to this linearized equation is the exponential decay
\begin{equation}
\label{wave-poor}
U(z)\sim e^{\alpha z}, \quad\qquad z\to-\infty.
\end{equation}
Substituting this form and (\ref{v}) into the linearized equation, the
decay constant $\alpha$ is the root of the following equation
\begin{equation}
\label{v-alpha}
1/2-r=\alpha^{-1}\left[(1-p)(1-e^{-\alpha})-r(e^{\alpha}-1)\right].
\end{equation}

Similarly, in the limit $z\to\infty$ we linearize the wave equation
(\ref{wave-eq}) for the small quantity $R=1-U$ to obtain
\begin{eqnarray*}
vR'=r[R(z)-R(z\!+\!1)]+p[R(z)-R(z\!-\!1)].
\end{eqnarray*}
In this case the behavior at large fitness is governed by the balance between
decline and advancement events where the fitter agent advances \cite{bkm}.
The solution to the above differential equation is again the exponential decay
\begin{equation}
\label{wave-rich}
R(z)\sim e^{-\beta z}, \quad\qquad z\to\infty,
\end{equation}
with the decay constant $\beta$ satisfying
\begin{equation}
\label{v-beta}
1/2-r=\beta^{-1}\left[p(e^{\beta}-1)-r(1-e^{-\beta})\right].
\end{equation}

\begin{figure}[t]
 \vspace*{0.cm}
\includegraphics*[width=0.4\textwidth]{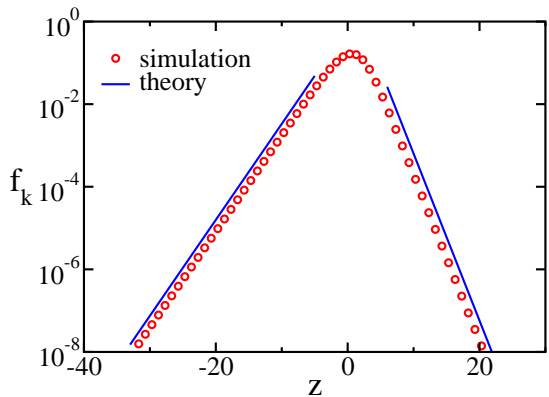}
\caption{Tails of the fitness distribution $f_k$ versus $z=k-vt$ for
  $p=r=1/4$. The theoretical predictions (\ref{wave-poor})--(\ref{v-beta})
  with $\alpha=0.535572$ and $\beta=0.930821$ are indicated by straight
  lines.}
\label{fig-tails}
\end{figure}

We conclude that the likelihood of having agents that are much richer
or much poorer than the average fitness $k=vt$ in the egalitarian
society is exponentially small, as illustrated in
Fig.~\ref{fig-tails}.  The society therefore consists of agents whose
fitnesses are all roughly the same, $k\approx vt$.  As one might
naturally anticipate, social inequalities are small under the dynamics
in which the rich preferentially gives to the poor.

\subsection{Less fit advances ($p=0$)}

For the case where the less fit agent always advances, the fitness
distribution has a special form. In this case, the complementary
cumulative distribution obeys
\begin{eqnarray}
\label{cum-eq2}
\frac{dG_k}{dt}&=&r(G_{k+1}-G_k)+\frac{1}{2}\left(G_{k-1}^2-G_k^2\right)
\end{eqnarray}
with the initial condition $G_k(0)=\delta_{k,-1}$ and the boundary
condition $G_{-1}(t)=1$.

\begin{figure}[t]
 \vspace*{0.cm}
\includegraphics*[width=0.4\textwidth]{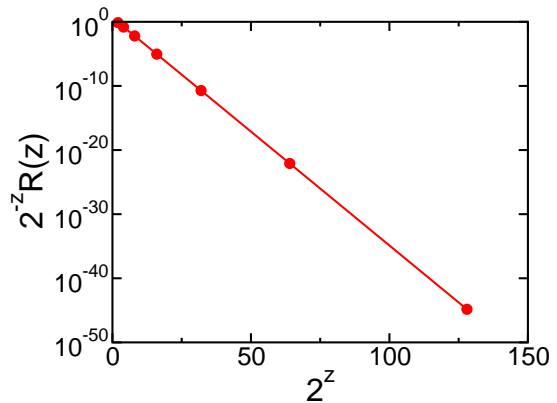}
\caption{The super-exponential tail: the quantity $2^{-z}R(z)$ versus
$2^z$, obtained by numerical integration of (\ref{cum-eq2}) to time
$t=100$ with $r=0$. }
\label{fig-se}
\end{figure}

We expect that the fitness distribution will continue to have the form
of a propagating wave.  Substituting the traveling wave form
$G_k(t)\to R(k-vt)$ into the master equation (\ref{cum-eq2}) gives
\begin{equation*}
-vR'(z)=r\left[R(z+1)-R(z)\right]+\frac{1}{2}\left[R^2(z-1)-R^2(z)\right].
\end{equation*}
An exponential solution does not give asymptotic balance of terms as
$z\to\infty$, and we therefore attempt a solution of the form
$R(z)\sim \psi(z)e^{-\phi(z)}$. Substituting this form into the
above equation and keeping only the dominant term
$\frac{1}{2}R^2(z-1)$ on the right-hand-side gives
\begin{equation}
\label{wave-eq-r}
v\psi(z)\phi'(z)e^{-\phi(z)}\approx \frac{1}{2}\psi^2(z-1)e^{-2\phi(z-1)}.
\end{equation}
For the positive terms on the left and the right hand side to balance,
the dominant exponential terms must first balance, yielding the
recursion equation $\phi(z)=2\phi(z-1)$. The solution is the
exponential $\phi(z)=C\,2^z$. Balancing the prefactors,
$v\psi(z)\phi'(z)=\frac{1}{2}\psi^2(z-1)$ yields $\psi(z)=8\ln 2Cv\,
2^z$.  As a result, the decay in the tail region is super-exponential
\cite{bkm-unpub}
\begin{equation}
\label{super-exp}
R(z)\sim 8\ln 2Cv\,2^z\exp\left(-C 2^z\right),
\end{equation}
as $z\to\infty$. The constants $C$ and $v$ should be determined numerically.
Hence, the front of the traveling wave is extremely sharp.  This tail
characterizes statistics of the rich, so when the rich never benefits from
interactions with the poor, rich agents are ultra-rare (Fig.~\ref{fig-se}).
Even though the leading tail extends to only a handful of sites, it is still
possible to verify the super-exponential decay (\ref{super-exp}).  In
contrast, the $z\to -\infty$ tail that characterizes the poor is not altered;
it has the same exponential tail as in Eq.~(\ref{wave-poor}).

\section{The Lower Class}
\label{poverty}

We now determine the fitness distribution of poorest agents, a class
that exists in both the hierarchical and lower-class societies.  As we
shall now show, the fitness distribution of the lower class in the
limit of small but non-zero fitness approaches a steady state.  For
the case of the hierarchical society, we write $F_k=L(1-g_k)$, where
$L$ is the lower-class fraction, with the deviation $g_k$ vanishing
for large $k$.  Substituting this form into the master equation
(\ref{cum-eq}) and setting the time derivative to zero, gives
\begin{equation}
r\frac{g_k-g_{k+1}}{g_{k-1}-g_k}=
1-p+(p-1/2)L\left(2-g_k-g_{k-1}\right).
\end{equation}

Consider first the lower-class society, for which the entire
population is poor, $L=1$.  Using this fact, and neglecting terms of
the order $g^2$, we find
\begin{equation}
\label{g-lc}
\frac{g_k-g_{k+1}}{g_{k-1}-g_k}=\frac{p}{r}.
\end{equation}
The solution to this equation is simply the exponential form $g_k\sim
\alpha^k$ with $\alpha=p/r$. Since $f_k=F_{k+1}-F_k= g_k-g_{k+1}$, then
\begin{equation}
\label{f-lc}
f_k\propto \left(\frac{p}{r}\right)^k.
\end{equation}
In the lower-class society, the fitness is confined to a very small
range.  Notice also that this exponential decay coincides with the
traveling wave solution (\ref{wave-rich}) with $v$ set equal to zero,
as the decay function is now $e^{-\beta}=p/r$.

Finally, we consider the hierarchical society.  Using (\ref{l}) and following
the same steps that led to Eq.~(\ref{g-lc}), we obtain
\begin{equation}
\frac{g_k-g_{k+1}}{g_{k-1}-g_k}=1-\gamma g_k
\end{equation}
with $\gamma=(r+p-1)/r$.  To determine $g_k$, we expand the differences
to second order and assume that $g''\ll g'$ to give, after straightforward
steps, $g''+\gamma gg'=0$. The asymptotic solution to this equation is
$g\simeq 2/(\gamma k)$.  Finally, using $f_k=F_{k+1}-F_k$ and $F_k=L(1-g_k)$,
we determine the fitness distribution from $f_k\simeq -Lg'$ to be
\begin{equation}
\label{fk-powerlaw}
f_k\simeq \frac{2r}{2p-1}\,k^{-2}.
\end{equation}
Thus, for the hierarchical society, the fitness distribution has a power-law
large-fitness tail in the lower class region (see also \cite{br} for more
details).

As discussed in section \ref{mobility}, there is a diffusive boundary
layer that separates the steady-state fitness distribution in the
lower class and time dependent fitness distribution in the middle
class.  From Eqs.~(\ref{scaling-form}) and (\ref{phi-hs}), the fitness
distribution in the middle class is $f_k\simeq
\left[(2p-1)t\right]^{-1}$.  Equating this expression with
Eq.~(\ref{fk-powerlaw}) gives a crossover scale
\begin{equation}
\label{crossover}
k_*\simeq \sqrt{2rt}.
\end{equation}
Thus, the steady-state region extends over a fitness range that grows
as $t^{1/2}$.  We also note that this crossover scale agrees with the
diffusivity $D_-=r$, obtained in section \ref{mobility}. In terms of
the variable $x=k/t$, the size of this region $x_*\sim t^{-1/2}$
decays with time.  Thus, a diffusive boundary layer separates the
lower class and the middle class.

\section{Conclusions}
\label{conc}

In summary, we have seen that the competition between advancement
and decline leads to a rich and realistic set of possible social
structures.  From the master equation for the underlying fitness
distribution, we obtain three types of classes: a static lower
class, a mobile but disperse middle class, and a mobile but
``condensed'' egalitarian class.  The population as a whole
organizes into four types of societies, three of which consist of
one of these classes, and a hierarchical society in which the lower
class and the middle class coexist.  Two parameters, the rate of
decline and the advancement probability, quantify the competition
between advancement and decline. The overall social organization is
determined solely by these two parameters.

The fitness distribution has a very different character in each of the
classes. In the lower class, this fitness distribution approaches a steady
state.  In the middle class, the distribution is self-similar in time and
correspondingly the characteristic fitness increases linearly with time.
Although agents are upwardly mobile, the disparities between agents in the
middle class also grows indefinitely.  In the egalitarian class, the fitness
distribution follows a traveling wave, so that all agents constantly advance,
but fitness differences between agents remains small.

Much of the richness of the phenomenology is due to the fact that the
mechanisms for advancement and decline are fundamentally different.
One requires interaction between agents, while the other is a
single-agent process.  This dichotomy is reflected by the master
equation where the decline terms are linear but the advancement terms
are nonlinear.  As a result, there is no detailed balance and the
dynamics are non-equilibrium in character.

It should be interesting to use the advancement-decline model to
analyze real-world data.  One natural application is to wealth and
income distributions of individuals, where both power-law and
exponential behavior has been observed \cite{ikr,dy}.  A related issue
is the wealth of nations.  It is well documented that the wealth
distribution of countries is extremely inequitable, with 60\% of the
world's population producing just 5.6\% of the planet's gross domestic
product (GDP), another 20\% producing 11.7\%, and the remaining 20\%
of the population producing 82.7\% of the GDP \cite{UN}.  The
existence of such a large underclass corresponds to a large decline
rate in our diversity model and it may be worthwhile to understand the
social mechanisms for such a large decline. Another possibility is
sports statistics where the winning percentage distribution of teams
plays the role of the fitness distribution \cite{bvr}.

\acknowledgments{We thank Philip Rosenau for useful discussions.  We
  acknowledge financial support from DOE grant W-7405-ENG-36 and
  NSF grant DMR0227670.}


\begin{thebibliography}{99}


\bibitem{idc} I. D. Chase, Amer.\ Sociological Rev.\ {\bf 45}, 905 (1980).

\bibitem{rvg} R. V. Gould, Amer.\ J. Sociology {\bf 107}, 1143 (2002).

\bibitem{hgl} H. G. Landau, Bull.\ Math.\ Biophys.\ {\bf 13}, 1 (1951).

\bibitem{eow} E. O. Wilson, {\it Sociobiology}, (Harvard University Press,
  Cambridge, MA, 1975).

\bibitem{eow1} E. O. Wilson, {\it The Insect Societies}, (Harvard University
  Press, Cambridge, MA, 1971).

\bibitem{wca} W. C. Allee, Biol.\ Symp.\ {\bf 8}, 139 (1942).

\bibitem{amg} A. M. Guhl, Anim.\ Behav.\ {\bf 16}, 219 (1968).

\bibitem{sf} M. W. Schein and M. H. Forman, Brit.\ J. Anim.\ Behav.\ {\bf 3}, 45 (1955).

\bibitem{vs} M. Varley and D. Symmes, Behaviour {\bf 27}, 54 (1966).

\bibitem{idc1} I. D. Chase, Behav.\ Sci.\ {\bf 19}, 374 (1980).

\bibitem{ikr}
  S.~Ispolatov, P.~L.~Krapivsky, and S.~Redner,
  Eur.\ Phys.\ Jour.\ B {\bf 2}, 267 (1998).

\bibitem{dy}
  A.~Dragulescu and V.~M.~Yakovenko,
  Eur.\ Phys.\ Jour.\ B {\bf 17}, 723 (2000).

\bibitem{wdan}
   G.~Weisbuch, G.~Deffuant, F.~Amblard, and J.~P.~Nadal,
   Complexity {\bf 7}, 55 (2002).

\bibitem{bkr} E.~Ben-Naim, P.~L.~Krapivsky, and S.~Redner, Physica D {\bf
    183}, 190 (2003).

\bibitem{smo} D.~Stauffer and H.~Meyer-Ortmanns, Int.\ J. Mod.\ Phys.\ B {\bf
    15}, 241 (2004).

\bibitem{wf} S.~Wasserman and K.~Faust, {\it Social Network Analysis}
  (Cambridge University Press, Cambridge, 1994).

\bibitem{gn} M.~Girvan and
  M.~E.~J.~Newman, Proc.\ Natl.\ Acad.\ Sci.\ USA {\bf 99}, 7821
  (2002).
\bibitem{ww} W.~Weidlich, {\it Sociodynamics: A Systematic Approach to
    Mathematical Modelling in the Social Sciences} (Harwood Academic
  Publishers, 2000)

\bibitem{vfh} D.~Helbing, I.~Farkas, and T.~Vicsek,
             Nature {\bf 407}, 487 (2000).

\bibitem{ckfl} I.~D.~Couzin, J.~Krause, N.~R.~Franks, S.~A.~Levin,
  Nature {\bf 433}, 513 (2005).


\bibitem{btd} E. Bonabeau, G. Theraulaz, and J.-L. Deneubourg, Physica A {\bf
    217}, 373 (1995).

\bibitem{SS} A. O. Sousa and D. Stauffer, Intl.\ J. Mod.\ Phys.\ C {\bf 5},
  1063 (2000); K. Malarz, D. Stauffer, and K. Kulakowski, 
  {\it physics/}0502118.

\bibitem{br} This special case was discussed in E. Ben-Naim and S. Redner,
             J. Stat. Mech L11002 (2005).

\bibitem{dz}
    D.~Zwillinger, {\it Handbook of Differential Equations}
    (Academic Press, London, 1989).

\bibitem{gbw}
    G.~B.~Whitham,
    {\it Linear and Nonlinear Waves},
    (Wiley, New York, 1974).

\bibitem{jmb} J. M. Burgers,
              {\it The nonlinear diffusion equation}
              (Reidel, Dordrecht, 1974).

\bibitem{bkm}
E.~Ben-Naim, P.~L.~Krapivsky, and S.~N.~Majumdar,
  Phys.\ Rev.\ E {\bf 64}, R035101 (2000).

\bibitem{bkm-unpub} E.~Ben-Naim, P.~L.~Krapivsky, and S.~N.~Majumdar,
unpublished.

\bibitem{UN} United Nations Development Program 1992, Human
  Development Report (Oxford University press for the United Nations
  Development Program, New York, 1992).

\bibitem{bvr} E.~Ben-Naim, F.~vazquez, and S.~Redner, ``What is the
   most competitive sport?'', preprint. 

\end{thebibliography}
\end{document}